\documentclass[journal]{IEEEtran}

\usepackage{amsmath,amssymb}
\usepackage{graphicx}
\usepackage{booktabs}
\usepackage{multirow}
\usepackage{siunitx}
\usepackage{subcaption}
\usepackage{xcolor}
\usepackage{url}
\usepackage[hidelinks]{hyperref}

\graphicspath{{figures/}{../thesis/figures/}}
\newcommand{\D}{\mathbf{D}}
\newcommand{\Dbase}{\mathbf{D}_{\mathrm{base}}}
\newcommand{\x}{\mathbf{x}}
\newcommand{\y}{\mathbf{y}}
\newcommand{\bDelta}{\boldsymbol{\Delta}}

\begin{document}

\title{LIPS-PC: Learning Illumination Patterns for Single-Pixel Classification}

\author{Tomer~Porat,~Talya Eden,~and~Eliahu Cohen%
\thanks{The authors are with Bar-Ilan University, Ramat-Gan 5290002, Israel.}%
\thanks{Submitted manuscript, \today.}}

\markboth{IEEE Transactions on Computational Imaging (submitted version)}%
{Porat \MakeLowercase{\textit{et al.}}: LIPS-PC for Single-Pixel Classification}

\maketitle

\begin{abstract}
In single-pixel (ghost) imaging, the object is probed by a sequence of known
illumination patterns and a detector with no spatial resolution returns one
number per pattern. Each pattern is a separate physical measurement, costing
acquisition time, photon budget and, for ionizing probes, delivered dose, so
the number of patterns $M$ is the quantity to minimize. A motivating instance
is automated radiographic defect screening on a production line, where
throughput and delivered dose are both set by the number of exposures taken
per item, and the decision needed is a class label rather than a
reconstructed image. We study whether the
pattern set should be \emph{learned} end-to-end for a downstream
\emph{classification} task instead of fixed to a generic transform, subjected to
a constraint that keeps the setting physically realizable: the learned
operator must be one fixed linear matrix, identical for every input. We
call this approach Learning Illumination Patterns for Single-Pixel
Classification (LIPS-PC). We parameterize it as a residual correction to a $2$D-DCT anchor,
$\D=\Dbase+\bDelta$, and train $\bDelta$ jointly with a compact classifier
under a cross-entropy classification objective, with no
image reconstructed at any point.
This is a task-driven linear embedding, obtained by optimization rather than
in closed form. We validate the approach first on two well-understood
digit/garment benchmarks (MNIST, Fashion-MNIST) under a controlled protocol,
then on the target application, a real industrial X-ray weld-radiograph
dataset for automated defect screening (RIAWELC). Across all three, over
five budgets $M\in\{8,16,32,64,128\}$ (sampling ratios from under
\SI{1}{\percent} to \SI{12.5}{\percent}), the learned operator beats the best
fixed transform by
$9$ to $15$ percentage points at $M=8$ on every dataset. Equivalently, at
$M=8$ it reaches an accuracy that the best fixed transform does not attain
until roughly $M=16$ to $M=32$, a $2\times$ to $4\times$ saving in
measurements at matched accuracy. The advantage generally shrinks as $M$
grows, and on RIAWELC it reverses by a small margin at intermediate budgets.
One of the main conclusions is that learning what to measure pays off especially
when measurements are very scarce.
\end{abstract}


\IEEEpeerreviewmaketitle

\section{Introduction}
\label{sec:intro}
\IEEEPARstart{A}{conventional} camera records millions of pixels in
parallel. A single-pixel camera~\cite{duarte2008}, and its ghost-imaging
(GI) and computational ghost-imaging (CGI)
relatives~\cite{pittman1995,shapiro2008,bromberg2009}, instead use one
detector with no spatial resolution, a so-called bucket detector. Spatial
information is recovered by illuminating the object with a sequence of known
patterns and correlating the single intensity readings with the patterns
that produced them. This paradigm is attractive, for instance, when a dense
sensor array is unavailable, unsafe, or expensive: imaging through
scattering media, imaging at wavelengths with no readily available focal-plane
array, and imaging with ionizing probes such as X-rays, where every photon
delivered to the object carries a dose cost. Here, single-pixel imaging motivates a
numerical sensing-design problem: which linear measurements best support classification?

\begin{figure}[t]
\centering
\includegraphics[width=\linewidth]{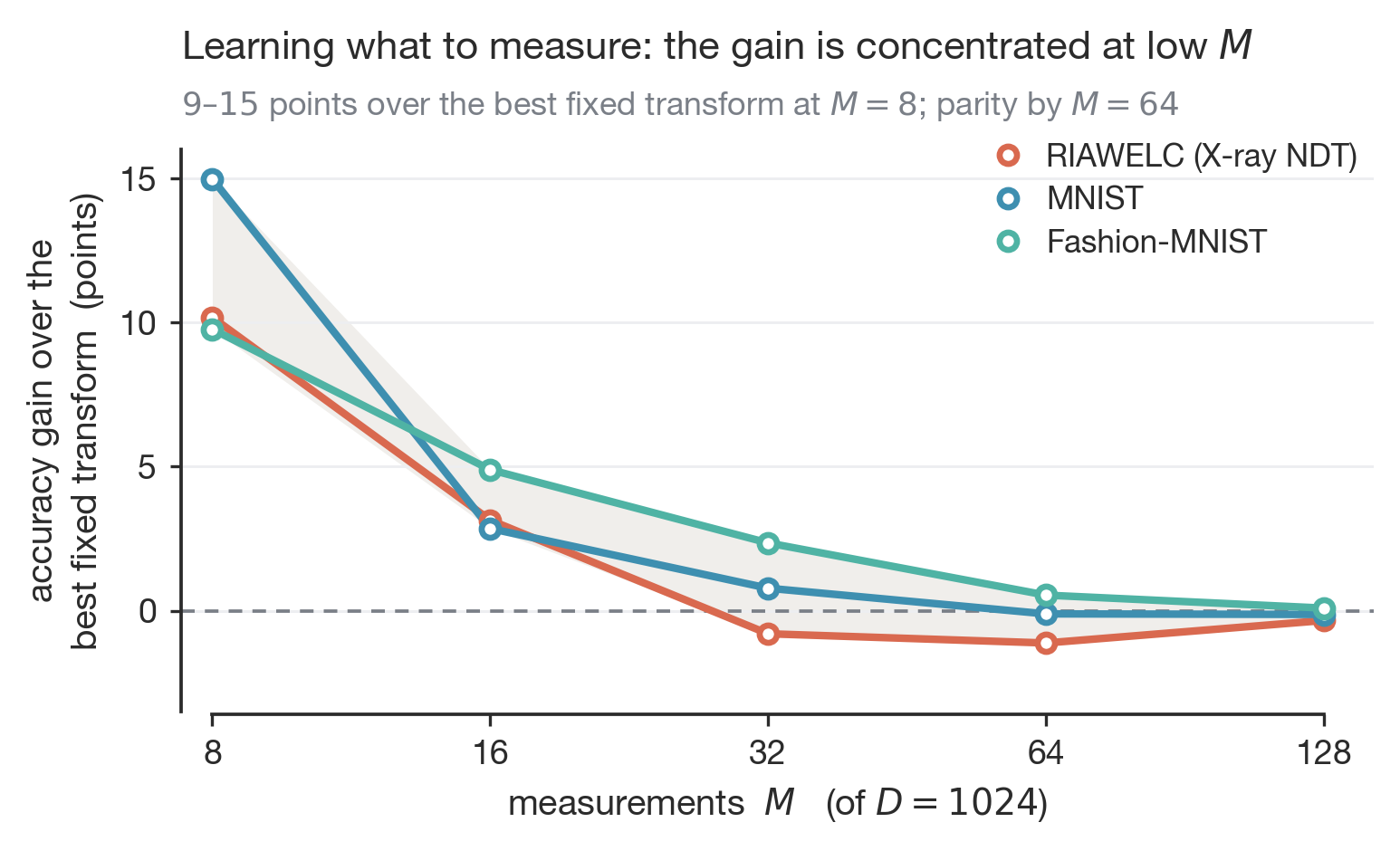}
\caption{Main result. Accuracy of the learned linear measurement operator
minus that of the best fixed transform, versus the number of measurements
$M$ (out of $D=1024$). On a real industrial X-ray dataset and two standard
benchmarks the learned operator wins by $9$ to $15$ percentage points at
$M=8$. The advantage shrinks to zero, or presents a small reversal, as $M$ grows.
Learning \emph{what} to measure helps most when there is very little budget
to spend.}
\label{fig:teaser}
\end{figure}

Beyond this physical motivation, the setting targets a concrete industrial need,
and it is the one this paper is ultimately built to serve. Radiographic defect
screening on a production line is currently limited by how many
exposures can be taken and processed per item; a measurement operator that reaches
high classification accuracy from very few readings offers a direct path to screening
more items per unit time, or at a lower delivered dose, using the same acquisition
hardware and without a human in the decision loop. The RIAWELC weld-defect case
studied below (Section~\ref{sec:setup}) is a concrete instance of this need: automatic,
fast defect identification rather than image reconstruction for its own sake. We reach
that case by first developing and validating the approach under tight experimental
control on MNIST and Fashion-MNIST, two digit/garment benchmarks whose class
structure and image statistics are well understood, where any failure of the method is
unambiguous; RIAWELC then tests whether the same, unmodified approach transfers to
noisier, lower-contrast real radiographs and a genuine downstream use case. The digit
benchmarks are a controlled stepping stone, not a second headline result alongside it.

In CGI the optical reference arm is removed and the patterns become
pre-defined, computer-generated masks applied directly to the illumination
path~\cite{shapiro2008,edgar2019}. The spatial resolution of the result is
then a property of the \emph{pattern set} rather than of any sensor, so the
pattern set becomes a design variable. Compressed
sensing~\cite{candes2006,donoho2006} explains why a small pattern set can
suffice: if the signal is sparse in some known basis and
suitable recovery conditions hold, far fewer than $D$
linear measurements can recover it.

Two observations motivate this work. First, most of the works that \emph{learn}
single-pixel patterns optimize them for reconstruction fidelity, using a
mean-squared-error or total-variation objective and a decoder that returns a
full image~\cite{higham2018,adler2016,lyu2017,wang2019}. Second, for many
end uses, such as sorting, inspection, and triage, the image is only a means
to a decision, and the decision may be reachable with far fewer
measurements than a faithful image would need. This has been shown for
biological cells~\cite{ota2018} and for coarse object
recognition~\cite{zhang2020,cao2021}. When the target is a class label, the
natural training objective is classification accuracy rather than pixel
fidelity, and the natural question is whether a task-optimized pattern set
beats a generic fixed transform at a fixed, small measurement budget.

Seen from the learning side, the object is this: train the linear
measurement operator jointly with the classifier that reads its output,
end-to-end, so that the few measurements taken are the ones most useful for
the decision. This is the reconstruction-free, task-driven end of
compressed learning~\cite{calderbank2009,adler2016cl,lu2026}, and the learned operator is a
supervised, task-driven linear embedding, a relative of supervised principal
component analysis and linear discriminant analysis, but reached by
gradient-based optimization under a hardware constraint rather than in
closed form.

We refer to this approach as Learning Illumination Patterns for Single-Pixel
Classification (LIPS-PC). We study this under one constraint that keeps the setting physically
realizable: the learned operator must be a single fixed linear matrix
$\D\in\mathbb{R}^{M\times D}$, identical for every object, applied as
$\y=\D\x$ before any nonlinear processing. A real bucket detector cannot
choose its illumination from the content of an object it has not yet sensed,
so per-image nonlinear ``encoders'' are excluded: such models reach high
accuracy but correspond to no realizable mask set. Within this constraint we
make three contributions.

\begin{enumerate}
\item A \emph{residual} parameterization of the learned pattern set,
$\D=\Dbase+\bDelta$, anchored to a $2$D discrete-cosine basis and trained
jointly with the classifier under a cross-entropy classification objective
(Section~\ref{sec:method}). The result is a task-aware linear embedding,
close in spirit to supervised principal component analysis, but obtained by
end-to-end optimization rather than in closed form.
\item An evaluation aimed at a real industrial X-ray non-destructive testing (NDT)
target, the weld-radiograph dataset RIAWELC~\cite{totino2023}, reached via a
controlled preliminary step on MNIST~\cite{lecun1998} and
Fashion-MNIST~\cite{xiao2017}, over five budgets spanning sub-one-percent
sampling ratios (Section~\ref{sec:setup}). To our knowledge this is the
first time task-learned single-pixel patterns are reported on a real
industrial radiographic dataset at this compression.
\item A consistent empirical finding (Fig.~\ref{fig:teaser}): the learned
advantage is largest at the smallest budget, between $9$ and $15$ percentage
points at $M=8$ on every dataset, and it generally decreases as $M$
grows larger towards $D$, with a small reversal on RIAWELC for $M\ge 32$
(Section~\ref{sec:results}). We report this as an observed pattern across
three datasets, consistent with compressed-sensing intuition, not as a
theorem.
\end{enumerate}

\section{Background and Related Work}
\label{sec:related}

\begin{figure}[t]
\centering
\includegraphics[width=0.92\linewidth]{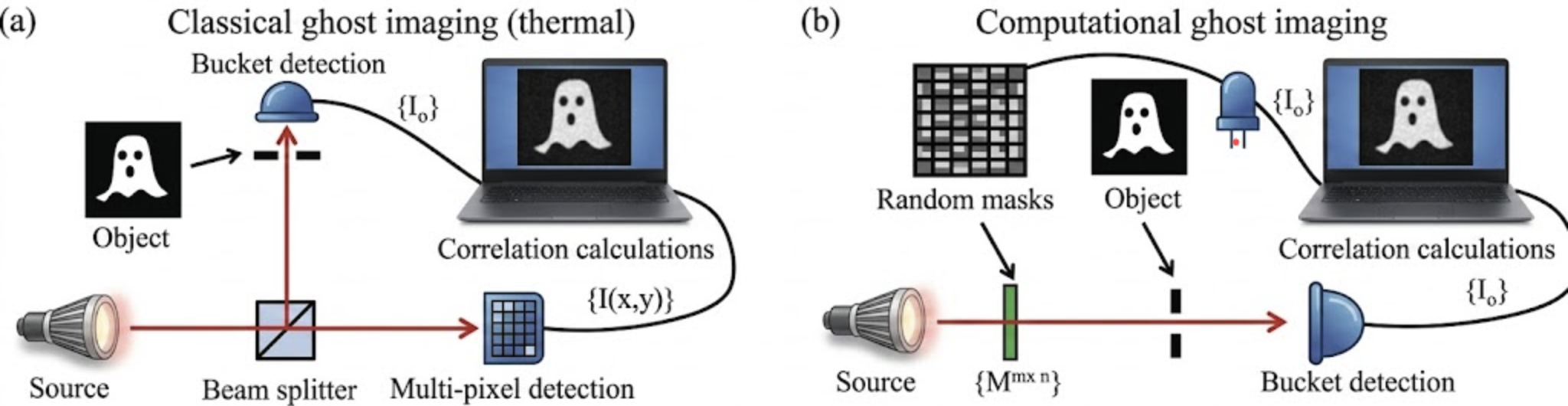}
\caption{Ghost-imaging architectures. Panel (a) shows a bucket detector in the
object arm and a spatially resolved detector in the reference arm. Panel
(b) shows computational ghost imaging, where known illumination patterns
replace the measured reference arm. The present numerical study uses the
ideal linear measurement model $\y=\D\x$.}
\label{fig:cgi}
\end{figure}

\subsection{Ghost imaging and CGI}
Ghost imaging originally used the correlation between a bucket detector in
the object arm and a spatially resolved detector in a correlated reference
arm~\cite{pittman1995,bromberg2009}. Computational ghost imaging removes the
reference arm and replaces it with known programmable patterns, for example
on a digital micromirror device~\cite{shapiro2008} (Fig.~\ref{fig:cgi}).
The measurement is then a deterministic linear map $\y=\D\x$, with $\D$ the
row-stack of vectorized patterns. Reviews of the modern single-pixel-imaging
pipeline, including pattern choice, detector noise, and reconstruction, are
given by~\cite{edgar2019}. The idea of computing on the compressed
measurements directly, rather than reconstructing first, goes back to
compressed learning~\cite{calderbank2009}, which showed that classifiers
trained in a random measurement domain retain much of their accuracy.

\subsection{Fixed pattern families}
Random patterns, drawn i.i.d.\ Gaussian or Bernoulli, are the
compressed-sensing baseline and satisfy the restricted-isometry property
with high probability~\cite{candes2006}. Hadamard patterns form an
orthogonal, binary-friendly basis; their sampling order matters, and
sequency (Walsh) ordering concentrates energy far better than the natural
order~\cite{sun2017}. The discrete cosine transform places most
natural-image energy in a few low-frequency coefficients and underlies JPEG.
All three families are generic in the sense that they are built around
signal structure, not around any particular downstream task.

\subsection{Learned patterns for reconstruction}
Deep networks have been used to reconstruct images from few single-pixel
measurements~\cite{higham2018,lyu2017,wang2019,kulkarni2016}, and to learn
the sensing matrix jointly with such a reconstruction
decoder~\cite{adler2016}. In all of these the training objective is image
fidelity. Closest to the present work, DO-CGI~\cite{hale2025} learns the
diffuser set itself, jointly with a Split-Bregman total-variation
reconstruction and a mean-squared-error loss, and reports clear gains over
random patterns at sampling ratios below five percent. Its objective is
still image fidelity. The present paper keeps the learned-diffuser idea but
replaces the reconstruction objective with a classification one and removes
the reconstruction step entirely.

\subsection{Learned patterns for classification}
Fewer works optimize the patterns for a decision. Calderbank et
al.~\cite{calderbank2009} give the theoretical basis. Ota et
al.~\cite{ota2018} classify cells image-free in a single-pixel flow
cytometer. Zhang et al.~\cite{zhang2020} classify fast-moving objects from
structured illumination without reconstruction. Cao et al.~\cite{cao2021}
report $94.23\%$ on handwritten digits from $M=16$ random single-pixel
patterns with a small network and no reconstruction step, a paradigm we
reproduce and exceed at $M=8$. Bacca et al.~\cite{bacca2019} jointly learn a
coded aperture and a classifier, including eight-measurement MNIST experiments at approximately $1\%$ sampling and a sweep over several sampling ratios. TransCL~\cite{mou2022} learns a linear
sensing layer that feeds a Transformer decoder and reaches about $91\%$ on
CIFAR-10, at $10\%$ sampling, and about $69\%$ at $1.8\%$ sampling; its decoder is
also far heavier than the minimal head used here. Closest in spirit, Han et
al.~\cite{han2026} study the analogous design problem for image-free
single-pixel \emph{segmentation}: how raw measurements are lifted into a
task representation, across a spectrum from fixed operators to
content-adaptive, nonlinear lifting networks, and report that this
adaptive lift is more robust to measurement noise than a
reconstruct-then-segment pipeline. Their strongest lift is content-adaptive
per input, which the physical-realizability constraint adopted here rules
out; our comparison instead stays within the fixed-operator end of that
spectrum (fixed transforms versus one jointly learned but still input-agnostic
matrix) and asks a classification-accuracy, not noise-robustness, question
across $M$. For context on the RIAWELC
task itself, full-resolution convolutional classifiers reach the high
nineties~\cite{totino2023,palma2024}, which confirms the four defect classes
are separable and frames the sub-Nyquist numbers below as a measurement
question, not a task-difficulty one.

\subsection{This work}
Relative to the above, the combination studied here is: (i) a
physical-realizability constraint on the learned operator, namely one fixed
linear matrix that is identical for every input; (ii) a classification
objective, not a reconstruction objective; and (iii) evaluation on a real
industrial X-ray NDT dataset at sampling ratios below one percent. Of these,
(iii) carries most of the paper's emphasis: MNIST and Fashion-MNIST serve as
an initial, controlled check of the approach on standard digit benchmarks,
before moving to the substantially more complex and practically relevant
setting of real industrial radiographs, which is where the approach is
intended to matter.

\section{Method}
\label{sec:method}

\begin{figure}[t]
\centering
\includegraphics[width=\linewidth]{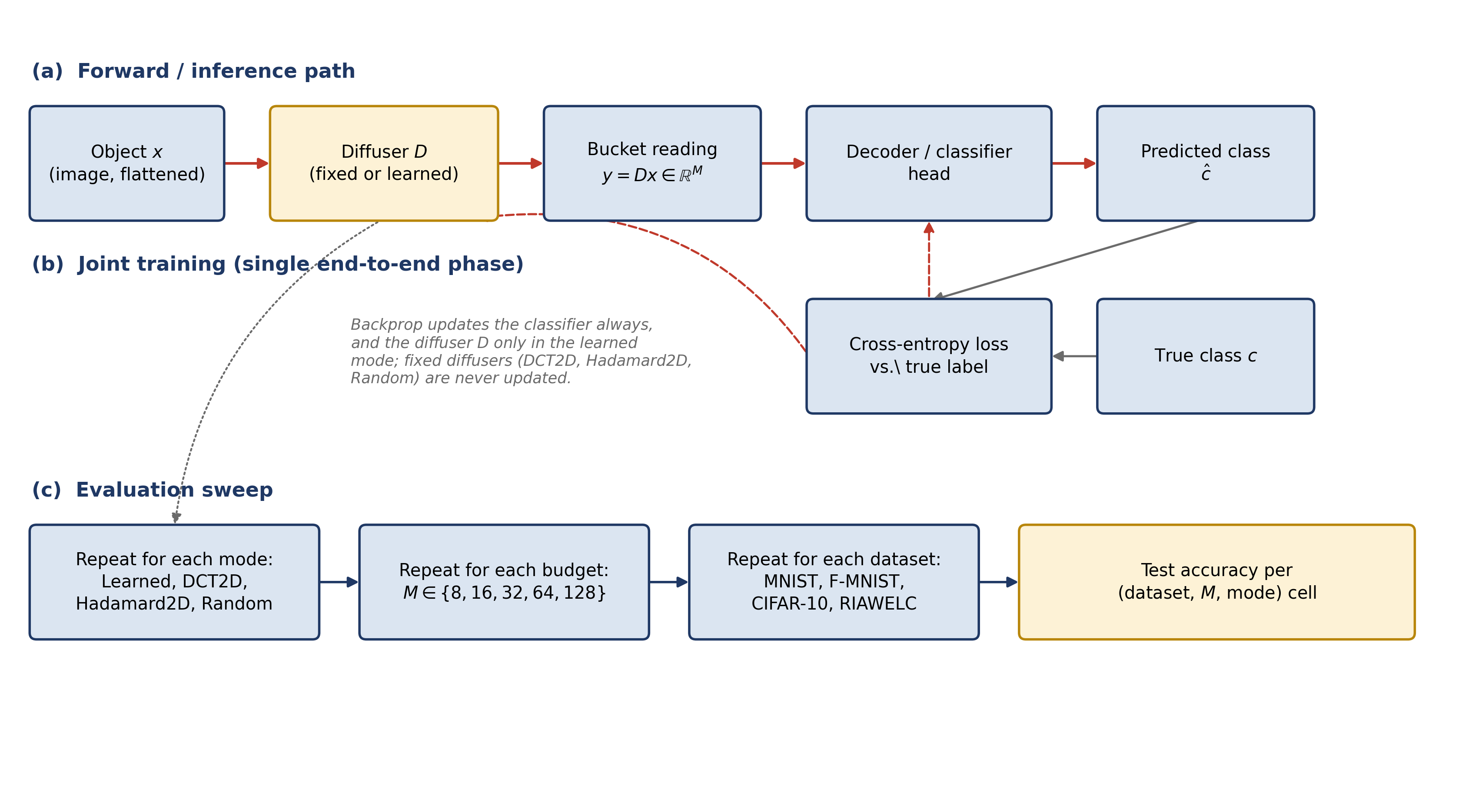}
\caption{Training and evaluation pipeline. An object $\x$ is compressed to
$\y=\D\x$ by a fixed $M\times D$ matrix, and a small decoder predicts the
class directly from $\y$, with no image reconstruction. The classifier is
always trained; $\D$ is trained only in the learned mode, jointly with the
classifier and by the cross-entropy loss alone setting the classification objective. The whole procedure is
repeated for every (diffuser mode, budget $M$, dataset) cell to produce
Table~\ref{tab:main}.}
\label{fig:pipeline}
\end{figure}

LIPS-PC comprises a small number of deliberate design choices. We state each
choice and its rationale in turn. Figure~\ref{fig:pipeline} shows how the
pieces fit together: the object is measured through a single fixed matrix
$\D$, the $M$-vector of readings is passed to a compact decoder, and, in the
learned mode only, gradients from the classification loss flow back into
$\D$ as well as into the decoder.

\subsection{Measurement model and non-adaptive sensing}
An image is a vector $\x\in\mathbb{R}^{D}$, with $D=1024$ for the
$32\times32$ inputs used here. We take $M$ measurements, each the inner
product of $\x$ with one fixed pattern. Stacking the patterns as the rows of
$\D\in\mathbb{R}^{M\times D}$ gives
\begin{equation}
\y=\D\,\x,\qquad \y\in\mathbb{R}^{M}.
\label{eq:model}
\end{equation}

\medskip\noindent\textbf{Constraint.}
$\D$ must be a single matrix, identical for every object, and linear. This
is exactly what a physical single-pixel mask set is: the detector integrates
$\x$ through a pattern and returns one number, before any nonlinear
computation. We therefore \emph{exclude} any nonlinear, per-image map that
would pose as a ``diffuser'', for example a convolutional network run on
$\x$. Such maps score highly but realize no mask set, so they are not
reported, not even as an ablation. The linear and input-agnostic form is
\emph{necessary but not sufficient} for realizability: a real device also
needs non-negative or sign-encoded patterns and finite bit depth. Those
issues are out of scope here and are discussed in
Section~\ref{sec:limitations}. They apply equally to the fixed baselines and
to the learned operator, so they do not affect the comparison.

\subsection{Fixed baselines}
The three fixed baselines span the standard options. Random is the
compressed-sensing baseline with no structure. Hadamard is a structured,
orthogonal, binary-friendly basis. DCT is the structured basis matched to
natural-image statistics. We use $2$D, per-channel variants and select the
$M$ lowest-frequency components in zigzag order for DCT, or the $M$
lowest-sequency components for Hadamard. All three are applied as genuine
near-orthonormal $M\times D$ matrices and are already
realizability-compliant in the linear-and-fixed sense above.

\subsection{Learned diffuser: a residual correction}
We do not learn $\D$ from scratch. An unconstrained matrix is often poorly
conditioned early in training and, in our experiments, less stable. Instead
we anchor to a strong basis and learn an additive correction:
\begin{equation}
\D \;=\; \Dbase + \bDelta ,
\label{eq:residual}
\end{equation}
where $\Dbase$ is the fixed $2$D-DCT basis and
$\bDelta\in\mathbb{R}^{M\times D}$ is a free tensor initialized to
$\mathbf{0}$, so that $\D=\Dbase$ exactly at the start of training. The
correction $\bDelta$ is optimized jointly with the classifier by the
classification cross-entropy loss; there is no
reconstruction term and no regularizer on $\bDelta$ beyond weight decay.
An unrestricted $\bDelta$ does not restrict the representable matrix family; the DCT anchor specifies the initialization. At inference $\D$ is a
single fixed matrix shared across all inputs, so Eq.~\eqref{eq:model} still
describes the a physically
realizable measurement.

Conceptually, $\D$ is a task-aware linear embedding. Like supervised
principal component analysis it is one global linear projection, but instead
of preserving variance it is optimized to preserve the directions that
separate the classes. This also predicts the behaviour observed in
Section~\ref{sec:results}: when $M$ is small the projection must be spent
economically, so a task-optimized choice of subspace matters a great deal;
as $M$ approaches $D$ almost any well-conditioned basis spans enough of the
signal, and the freedom to trient $\D$ buys little.

\subsection{Decoder as a control}
\begin{figure}[t]
\centering
\includegraphics[width=\linewidth]{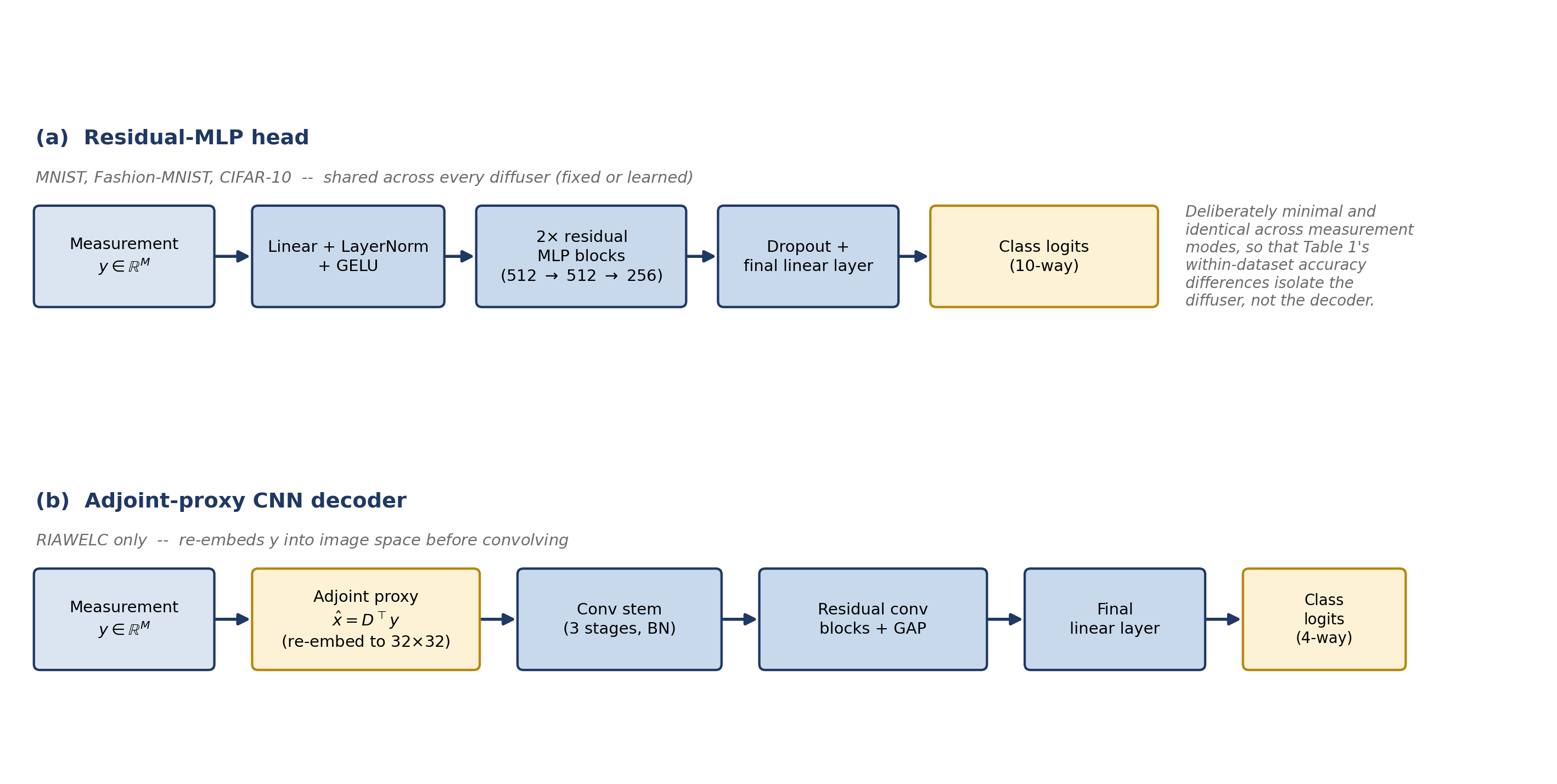}
\caption{Decoder heads. Panel (a): for MNIST and
Fashion-MNIST, every sensing mode
feeds the same residual multilayer perceptron on the raw $M$-vector $\y$.
Panel (b): for RIAWELC, the measurements are reembedded
as an adjoint proxy
image $\hat{\x}=\D^{\top}\y$ and classified by a small convolutional
network; all RIAWELC modes share this head.}
\label{fig:arch}
\end{figure}

The same classifier architecture is used for every pattern set, so that a difference in
accuracy can be attributed to the measurement operator rather than to
decoder capacity. For MNIST and Fashion-MNIST every mode, the three fixed
baselines and the learned one, feeds an identical residual multilayer
perceptron (Fig.~\ref{fig:arch}, left): a linear projection of $\y$ to width
$512$, three residual blocks at width $512$ with batch normalization and
dropout, and a $512\!\to\!256\!\to\!C$ output head. The learned mode adds
only the $M\times D$ parameters of $\bDelta$ on top of this shared head. Holding the decoder architecture fixed is what makes the comparison a
comparison of measurement operators: any change in accuracy between modes is
then a property of $\D$, since nothing downstream of $\y$ has changed.

\subsection{The RIAWELC variant}
\begin{figure}[t]
\centering
\includegraphics[width=\linewidth]{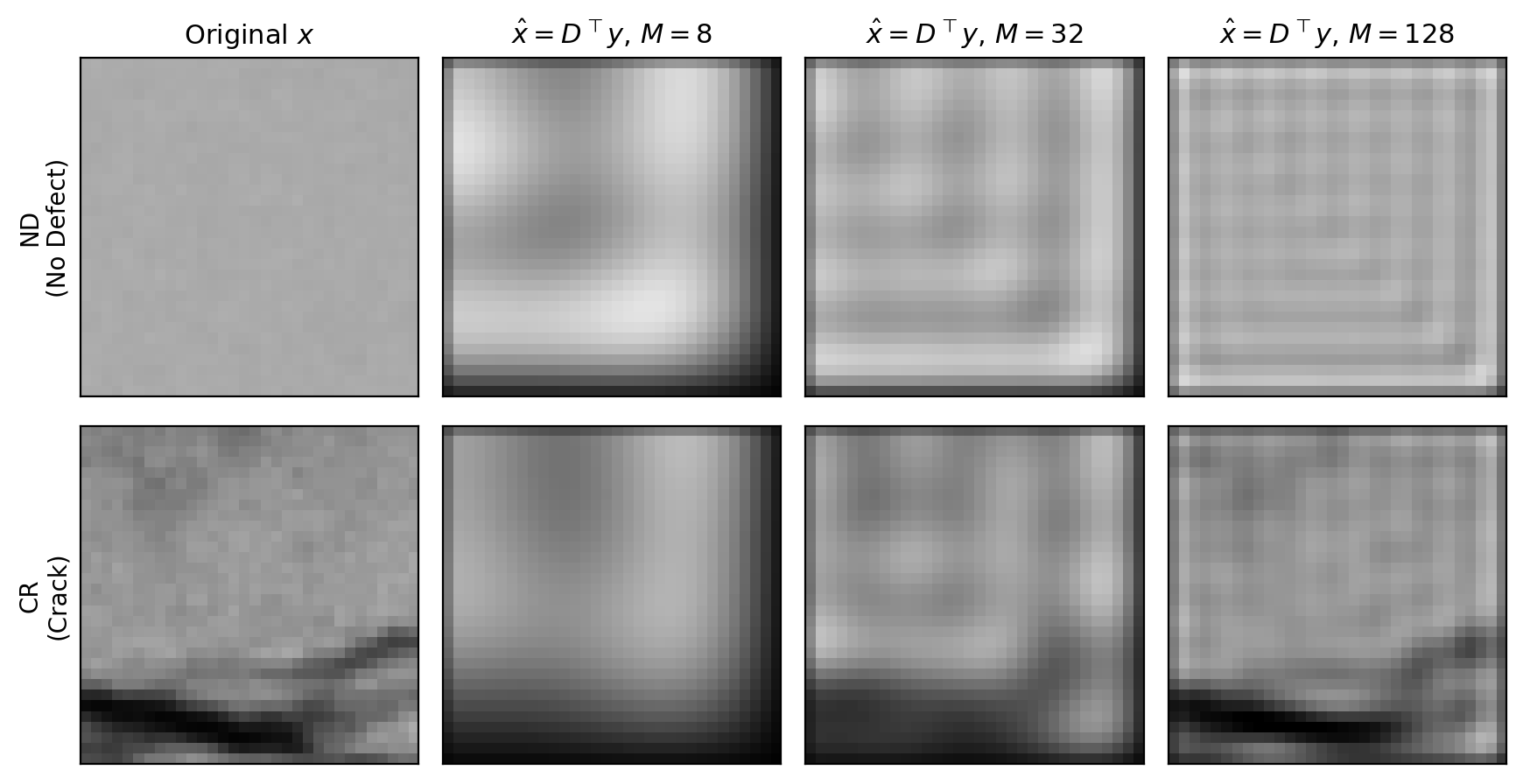}
\caption{The adjoint-proxy re-embedding $\hat{\x}=\D^{\top}\y$ on real
RIAWELC crops, for two of the four defect classes (no defect, crack; lack of
penetration and porosity follow the same trend and are omitted here for
space, code in Section~\ref{sec:recipe}), illustrated with the fixed
$2$D-DCT anchor $\D=\Dbase$ at $M=8$, $32$, and $128$ (not the learned
matrix, so this shows the decoder's re-embedding mechanism in isolation from
the learned correction). Resolution in the proxy increases with $M$; this is
a visual illustration of the mechanism, not a measurement of accuracy.}
\label{fig:proxy}
\end{figure}

RIAWELC is treated differently, deliberately, because its effective training
set is small: it contains thousands of image crops but they are drawn from
only about $29$ physical welds. A free $M\times D$ tensor $\bDelta$ then has
enough capacity to fit weld-specific nuisance structure. We therefore
replace it with a \emph{network-generated} correction,
$\bDelta=f_\theta(\mathbf{z})$, where $f_\theta$ is a small shared
network applied to a fixed latent vector $\mathbf{z}$; this gives the
correction far fewer effective degrees of freedom. The RIAWELC decoder is
also different: rather than a multilayer perceptron on $\y$, we form the
adjoint proxy image $\hat{\x}=\D^{\top}\y$ and classify it with a compact
three-stage residual convolutional network ($64\text{-}128\text{-}256$
channels, global average pooling, followed by one linear layer that maps
the pooled features to class logits). The map
$\y\mapsto\D^{\top}\y$ is linear and adds no information; it only re-embeds
the $M$ measurements into image coordinates so that a convolutional decoder
can exploit spatial structure~\cite{adler2016cl}. All RIAWELC modes share this
same decoder, so the comparison inside RIAWELC remains controlled.
Figure~\ref{fig:proxy} shows what this re-embedding looks like on actual weld
radiographs, for two representative defect classes, rather than only in
schematic form (Fig.~\ref{fig:arch}): at $M=8$ the proxy is a coarse blur
that keeps only gross shape, and by $M=128$ class-relevant structure, such
as the crack in the second row, is already visible, consistent with the
accuracy trend of Table~\ref{tab:main}.

\medskip\noindent\textbf{A failure mode, and its fix.}
A shared generator couples every entry of $\bDelta$ to each weight update.
In one run we observed the correction norm diverging,
$\lVert f_\theta(\mathbf{z})\rVert\approx 4400$ against
$\lVert\Dbase\rVert\approx 3$, that is, the anchor was effectively
discarded. Two changes, both needed, remove this behaviour: a separate and
lower learning rate for $f_\theta$, and a hard structural cap
$\lVert f_\theta(\mathbf{z})\rVert\le c\,\lVert\Dbase\rVert$ enforced by
construction (we rescale the generator output whenever it exceeds the cap).
The point is transferable to any design in which a network produces a
correction to a fixed operator.

\subsection{Training protocol and recipe}
\label{sec:recipe}
For a fixed baseline, training is a single stage: a classifier is trained on
the fixed measurements. For the learned mode there are two stages. In
\emph{stage~1} the diffuser and a classifier are trained together,
end-to-end, by the classification cross-entropy loss; this is where
$\bDelta$ is learned. In \emph{stage~2} the learned matrix $\D$ is frozen
and a \emph{fresh} classifier is trained on it, exactly as for a fixed
baseline. The number reported for every mode, learned or fixed, is the
stage-2 (or single-stage) test accuracy, so all four modes are scored under
an identical classifier-training protocol and only the measurement operator
differs.\footnote{Code: \url{https://github.com/analytics972/lips-pc}.}

All stages use stochastic gradient descent with Nesterov momentum $0.9$, a
five-epoch linear warm-up followed by cosine decay, mixup ($\alpha=0.2$) on
the measurement vectors, label smoothing $0.1$, gradient-norm clipping at
$1.0$, and weight decay $5\times10^{-4}$. The batch size is $64$. Stage~1
uses a learning rate of $1\times10^{-3}$ for $\bDelta$ (and for the RIAWELC
generator $f_\theta$) and $5\times10^{-3}$ for its classifier; the lower
rate on $f_\theta$, together with the norm cap $c=2$ of
Section~\ref{sec:method}, is what keeps the generated correction from
diverging. The stage-2 and fixed-baseline classifiers use a peak learning
rate of $0.1$. Each stage runs for $10$ epochs on MNIST and Fashion-MNIST
and for $35$ epochs on RIAWELC, which converges more slowly. Table
\ref{tab:hparams} collects these values. Every
(dataset, $M$, mode) combination is a single training run under one fixed
random seed; the statistical consequence is discussed in
Section~\ref{sec:limitations}.

\begin{table}[t]
\centering
\caption{Hyperparameters and module sizes.}
\label{tab:hparams}
\footnotesize
\setlength{\tabcolsep}{4pt}
\begin{tabular}{@{}l p{0.60\columnwidth}@{}}
\toprule
Setting & Value \\
\midrule
Input size / $D$        & $32\times32$ / $1024$ \\
Optimizer               & SGD, Nesterov, momentum $0.9$ \\
Epochs per stage        & $10$ (MNIST, F-MNIST); $35$ (RIAWELC) \\
LR schedule             & $5$-epoch warm-up, then cosine decay \\
Stage-1 LR              & $1\times10^{-3}$ ($\bDelta$, $f_\theta$); $5\times10^{-3}$ (classifier) \\
Stage-2 / baseline LR   & $0.1$ \\
Batch size              & $64$ \\
Weight decay            & $5\times10^{-4}$ \\
Mixup $\alpha$          & $0.2$ \\
Label smoothing         & $0.1$ \\
Gradient-norm clip      & $1.0$ \\
Grayscale decoder head     & residual MLP, width $512$, $3$ blocks, then $512\!\to\!256\!\to\!C$ \\
Block dropout           & $0.15$ (blocks), $0.1$ (output) \\
RIAWELC decoder         & $3$-stage residual CNN, $64\text{-}128\text{-}256$ ch, GAP + linear \\
RIAWELC $f_\theta$      & MLP, fixed latent $100$, hidden $512$; norm cap $c=2.0$ \\
\bottomrule
\end{tabular}
\end{table}

\section{Experimental Setup}
\label{sec:setup}

\subsection{Datasets}
\begin{figure*}[t]
\centering
\includegraphics[width=0.95\linewidth]{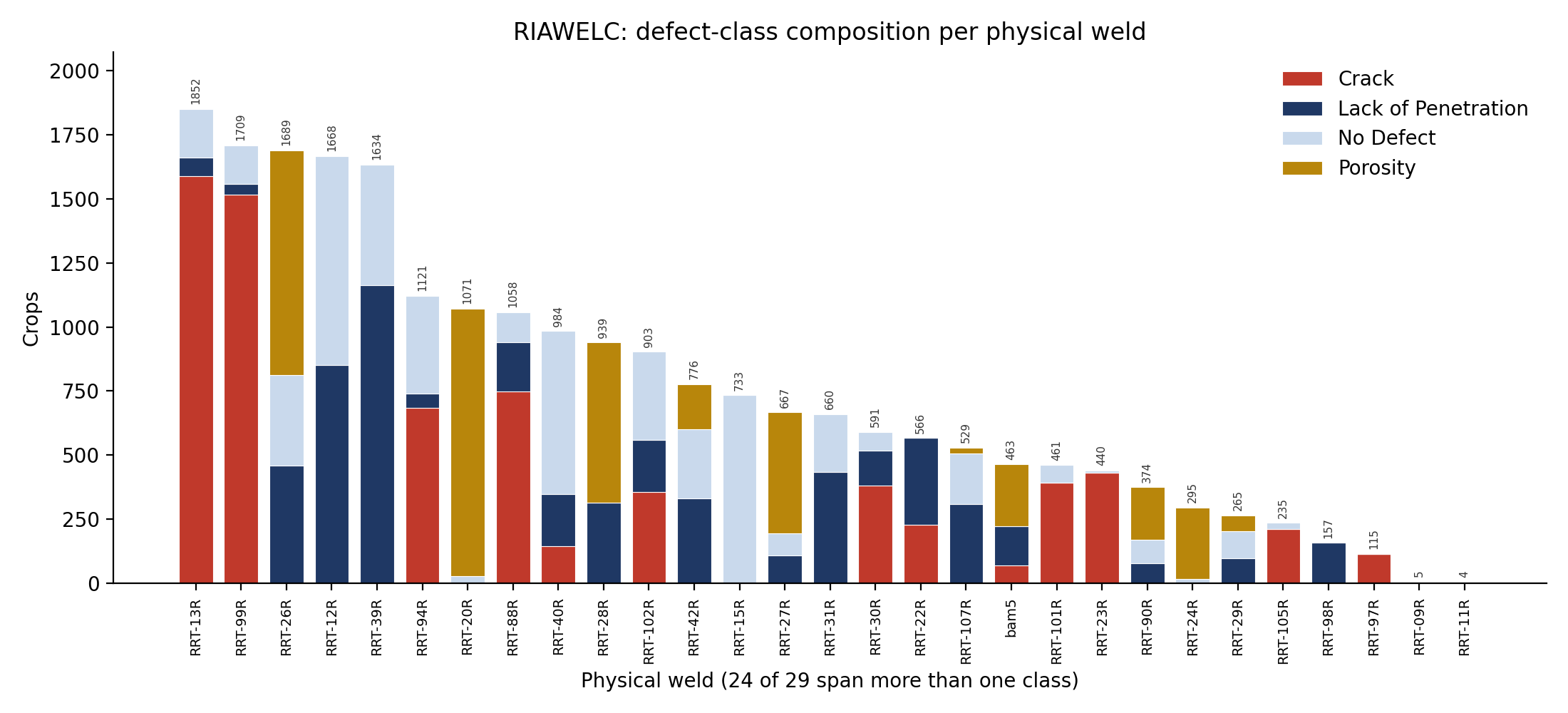}
\caption{Defect-class composition of every physical weld in RIAWELC, sorted
by crop count. Most welds (24 of 29) span more than one defect class, so
weld identity alone cannot predict the label under the crop-level split
used here.}
\label{fig:weldcomp}
\end{figure*}

\textbf{MNIST}~\cite{lecun1998} and \textbf{Fashion-MNIST}~\cite{xiao2017}
are $28\times28$ grayscale, ten classes, resized to $32\times32$ for a
common input size. They serve as the controlled preliminary step described
in Section~\ref{sec:intro}: well-characterized class structure against
which the method and its failure modes are easy to read, before testing
transfer to the real target. \textbf{RIAWELC}~\cite{totino2023}, the target
application, contains real industrial X-ray weld radiographs in four
classes: no defect, crack, lack of penetration, and porosity. Images are
natively $224\times224$ and are resized to $32\times32$. Low contrast and
high background noise make this dataset qualitatively different from the
digit benchmarks. We use an $80/20$ split, stratified by class, at the level
of individual crops. Because the crops of a given weld are visually
correlated, we checked directly whether weld identity could act as a label
shortcut under this split, by tabulating the defect-class composition of
every physical weld (Fig.~\ref{fig:weldcomp}). It could not: $24$ of the
$29$ welds ($82.8\%$) contribute crops to more than one defect class, so
knowing which weld a crop came from does not determine its label. This
rules out that specific shortcut; it does not rule out near-duplicate crops
on either side of the split as a separate channel for inflated accuracy,
which is a residual risk discussed in Section~\ref{sec:limitations} and
not resolved by this check. CIFAR-10
and other natural-image data are outside the scope of this paper and are
left to future work, since the RGB, high-intra-class-variance regime
behaves differently and deserves its own treatment.

\subsection{Budgets}
We use $M\in\{8,16,32,64,128\}$ out of $D=1024$, that is sampling ratios
from $0.78\%$ to $12.5\%$, entirely in the sub-Nyquist regime. The smallest
budget, $M=8$, is the operating point of most interest for dose- or
time-limited acquisition.

\subsection{Compute}
All experiments run on a single workstation: an Apple M4 with $24$\,GB of
unified memory and no discrete GPU, training on the integrated GPU through
the MPS backend. This modest budget is the reason for the lightweight
decoder heads and for the single-run-per-cell protocol.

\section{Results}
\label{sec:results}

\subsection{Central message}
\begin{figure*}[t]
\centering
\begin{subfigure}{0.325\linewidth}
  \includegraphics[width=\linewidth]{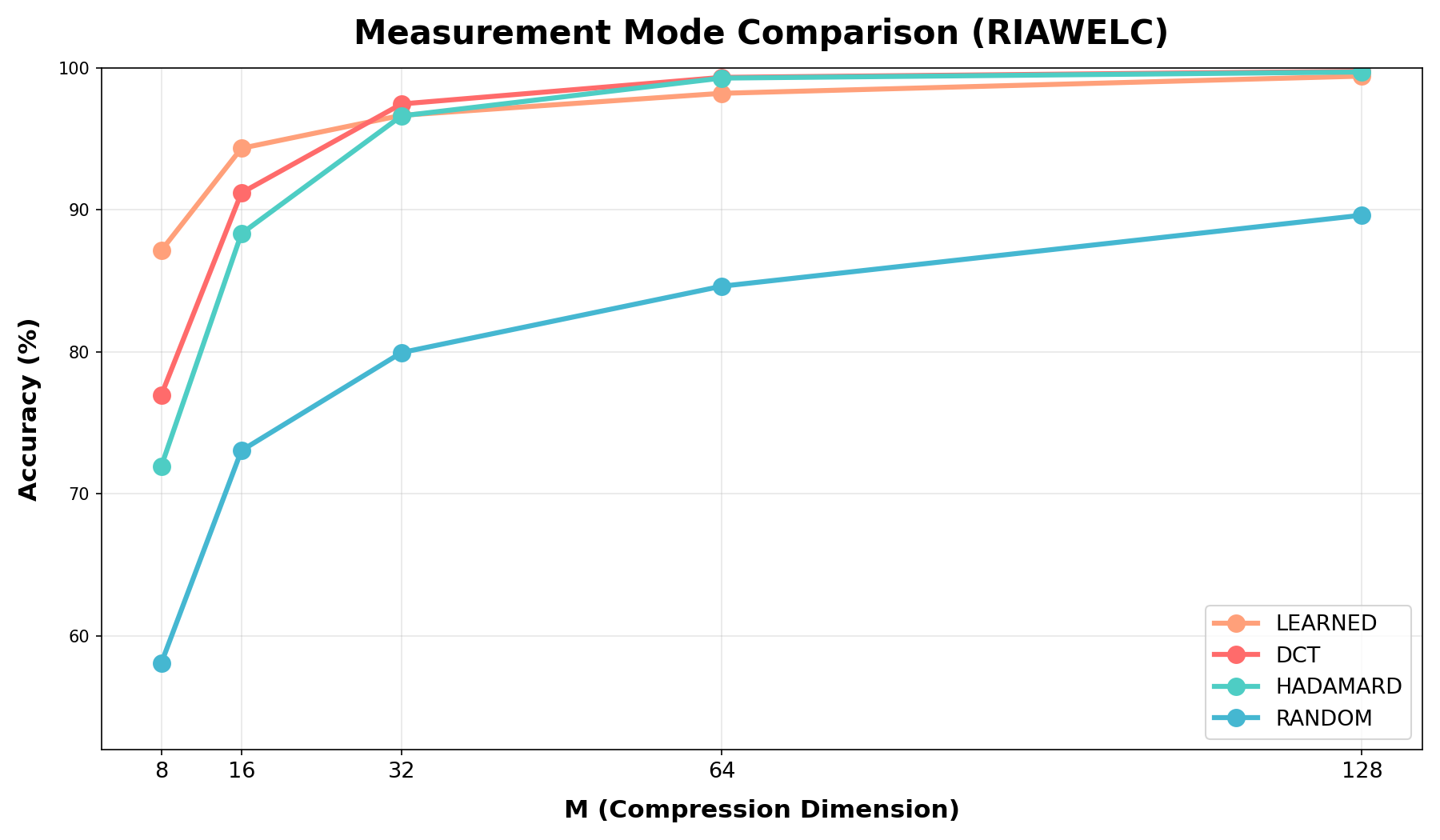}
  \caption{RIAWELC (X-ray weld radiographs)}
\end{subfigure}\hfill
\begin{subfigure}{0.325\linewidth}
  \includegraphics[width=\linewidth]{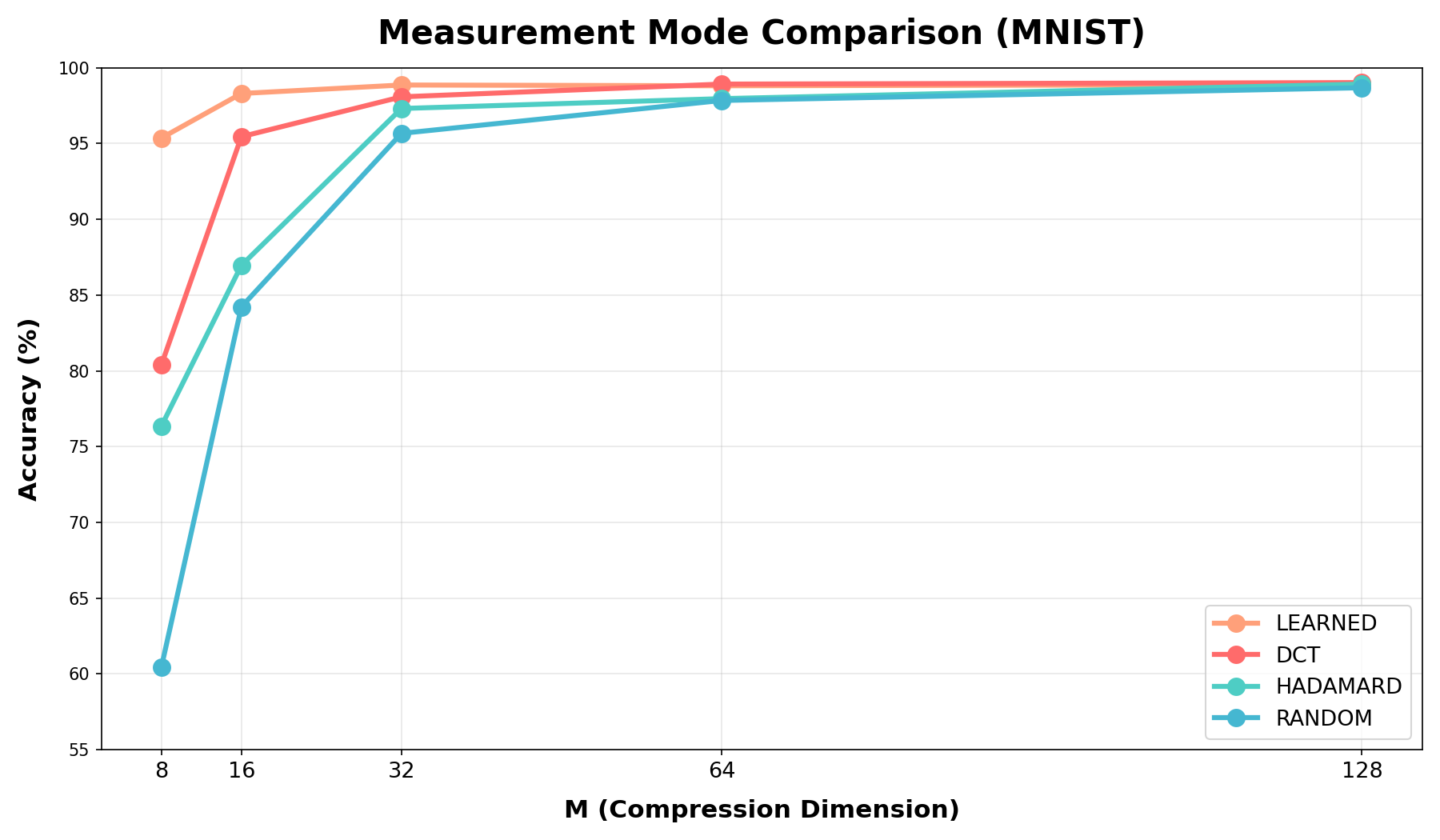}
  \caption{MNIST}
\end{subfigure}\hfill
\begin{subfigure}{0.325\linewidth}
  \includegraphics[width=\linewidth]{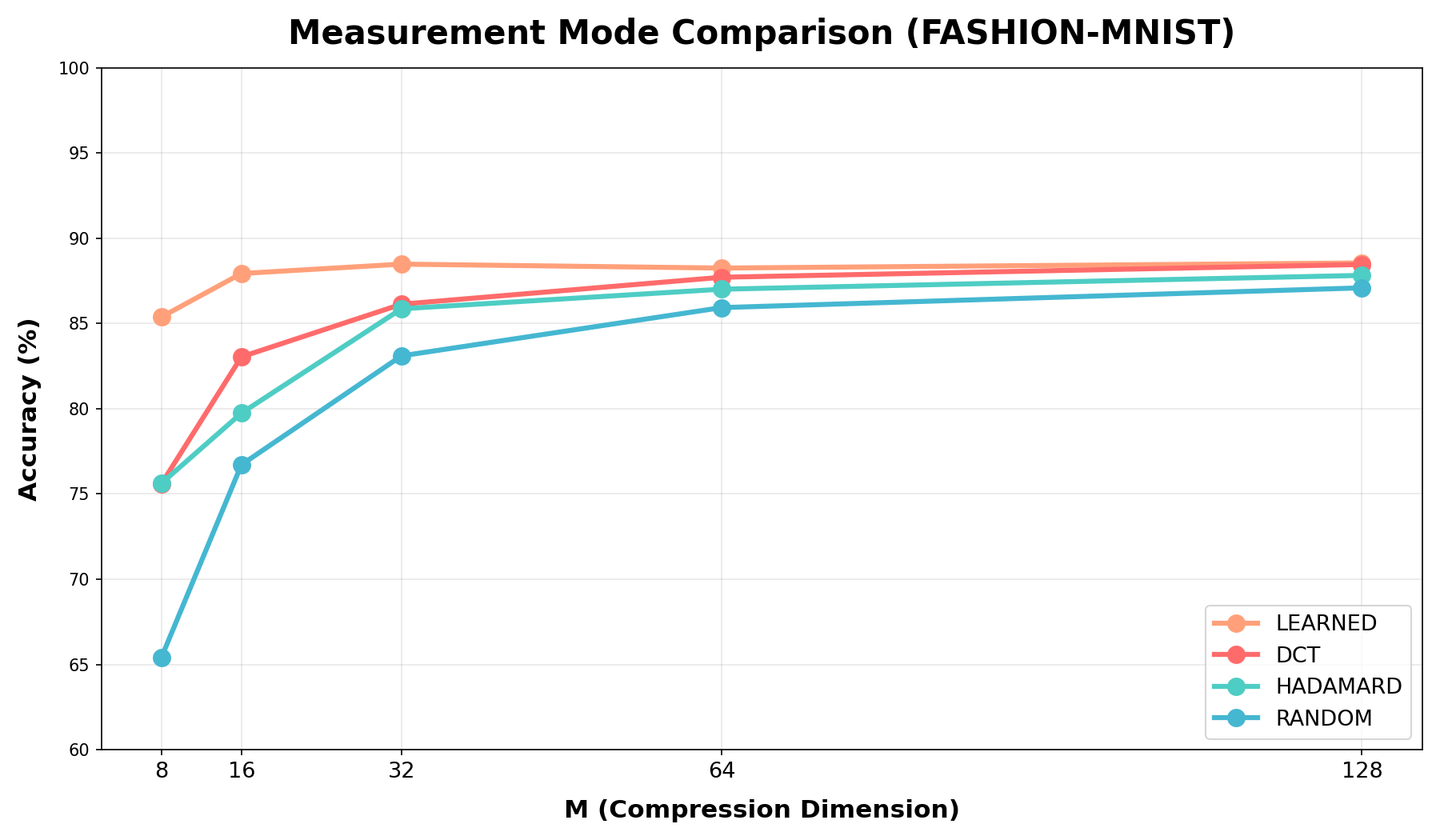}
  \caption{Fashion-MNIST}
\end{subfigure}
\caption{Test accuracy versus measurement budget $M$: learned diffuser
against the strongest fixed transform (DCT2D) and against random Gaussian
patterns. On every dataset the learned curve is well above the fixed curves
at $M=8$ and the gap closes as $M$ grows. On RIAWELC the learned curve leads
at $M=8$ and $M=16$, is overtaken from $M=32$, and the deficit widens before
narrowing again by $M=128$. MNIST and Fashion-MNIST use the free-tensor
correction; RIAWELC uses the network-generated correction and CNN decoder of
Section~\ref{sec:method}.}
\label{fig:acc}
\end{figure*}

On all three datasets the learned pattern set's advantage over the best
fixed transform is concentrated at the smallest budgets and decays as $M$
grows. Figure~\ref{fig:acc} plots test accuracy against $M$. The fixed
baseline that wins among the three is always DCT2D, so the figure shows the
learned mode against DCT2D and against Random; the full four-way numbers are
in Table~\ref{tab:main}.

\subsection{Full numbers}
Table~\ref{tab:main} gives test accuracy for all four modes and all five
budgets. At $M=8$ LIPS-PC is the best mode on every dataset.
Random is almost always last, so the substantive comparison is learned
versus DCT2D. On MNIST the learned mode stays best or within $0.13$
percentage points of best at every budget. On Fashion-MNIST it is best at
every budget. On RIAWELC it is far ahead at $M=8$ and $M=16$, then is
overtaken by DCT2D and Hadamard2D from $M=32$ onward, though by a small
margin that never exceeds about one point.

\begin{table}[t]
\centering
\caption{Test accuracy (\%). \textbf{Bold} = best mode at that budget.
MNIST and Fashion-MNIST use the free-tensor residual correction and the MLP
head; RIAWELC uses the network-generated correction and the CNN decoder
(Section~\ref{sec:method}). ``Hadam.''\ = Hadamard2D.}
\label{tab:main}
\small
\begin{tabular}{@{}ll rrrr@{}}
\toprule
Dataset & $M$ & Learned & DCT2D & Hadam.\ & Random \\
\midrule
\multirow{5}{*}{RIAWELC}
 & 8   & \textbf{87.14} & 76.97 & 71.96 & 58.10 \\
 & 16  & \textbf{94.33} & 91.19 & 88.32 & 73.05 \\
 & 32  & 96.65 & \textbf{97.45} & 96.61 & 79.95 \\
 & 64  & 98.20 & \textbf{99.32} & 99.27 & 84.62 \\
 & 128 & 99.41 & \textbf{99.75} & 99.70 & 89.62 \\
\midrule
\multirow{5}{*}{MNIST}
 & 8   & \textbf{95.36} & 80.39 & 76.35 & 60.44 \\
 & 16  & \textbf{98.30} & 95.45 & 86.96 & 84.22 \\
 & 32  & \textbf{98.86} & 98.08 & 97.31 & 95.67 \\
 & 64  & 98.81 & \textbf{98.92} & 97.96 & 97.84 \\
 & 128 & 98.89 & \textbf{99.02} & 98.92 & 98.68 \\
\midrule
\multirow{5}{*}{F-MNIST}
 & 8   & \textbf{85.38} & 75.59 & 75.62 & 65.40 \\
 & 16  & \textbf{87.93} & 83.04 & 79.74 & 76.71 \\
 & 32  & \textbf{88.48} & 86.13 & 85.86 & 83.10 \\
 & 64  & \textbf{88.25} & 87.71 & 87.01 & 85.93 \\
 & 128 & \textbf{88.55} & 88.46 & 87.82 & 87.09 \\
\bottomrule
\end{tabular}
\end{table}

\subsection{The margin, and how it scales}
Table~\ref{tab:margins} isolates the learned diffuser's margin over the best
fixed transform at each budget, and Fig.~\ref{fig:teaser} plots it. The
margin is between $9.76$ and $14.97$ percentage points at $M=8$ on every
dataset, and it generally shrinks thereafter. On MNIST and Fashion-MNIST the decay is mild and near-monotone,
and the margin stays at or above zero except for a dip of at most $0.13$
points on MNIST at the two largest budgets. On RIAWELC the margin turns
negative earlier, at $M=32$, and non-monotonically: the largest deficit,
$-1.12$ points, is at $M=64$, and it narrows again to $-0.34$ points at
$M=128$. This is what the embedding view in Section~\ref{sec:method}
predicts: the value of a task-optimized subspace is largest when the budget
is tightest. We present it as an empirical pattern across three datasets,
not as a proof, and note that the RIAWELC crossover in particular rests on
single runs (Section~\ref{sec:limitations}).

The same numbers read as a measurement saving. The learned diffuser's $M=8$
accuracy is one that the best fixed transform does not reach until $M=16$ on
MNIST ($95.36$ vs.\ $95.45$ at $M=16$) and on RIAWELC ($87.14$, between the
fixed $M=8$ and $M=16$ values), and until roughly $M=32$ on Fashion-MNIST
($85.38$, between the fixed $M=16$ and $M=32$ values). At the small-budget
operating point that matters for dose- or time-limited acquisition, the
learned operator therefore enables $2\times$ to $4\times$ reduction in the
number of physical measurements at matched accuracy.

\begin{table}[t]
\centering
\caption{Learned-diffuser margin over the best fixed transform (percentage
points). Positive = learned wins.}
\label{tab:margins}
\small
\begin{tabular}{@{}l rrrrr@{}}
\toprule
Dataset & $M{=}8$ & $M{=}16$ & $M{=}32$ & $M{=}64$ & $M{=}128$ \\
\midrule
RIAWELC        & $+10.17$ & $+3.14$ & $-0.80$ & $-1.12$ & $-0.34$ \\
MNIST          & $+14.97$ & $+2.85$ & $+0.78$ & $-0.11$ & $-0.13$ \\
F-MNIST        & $+9.76$  & $+4.89$ & $+2.35$ & $+0.54$ & $+0.09$ \\
\bottomrule
\end{tabular}
\end{table}

\subsection{Comparison with a published single-pixel classifier}
Cao et al.~\cite{cao2021} report $94.23\%$ on handwritten digits from $M=16$
random single-pixel patterns. On MNIST the learned diffuser here reaches
$95.36\%$ at $M=8$, that is, a comparable accuracy at half the measurement
budget. The two settings differ in dataset preprocessing, network, and
training protocol, so this is a point of context rather than a controlled
head-to-head, but it places the numbers here in the range of prior
reconstruction-free single-pixel classification while using fewer
measurements.

\subsection{What the learned patterns look like}
\begin{figure*}[t]
\centering
\begin{subfigure}{0.30\linewidth}
  \includegraphics[width=\linewidth]{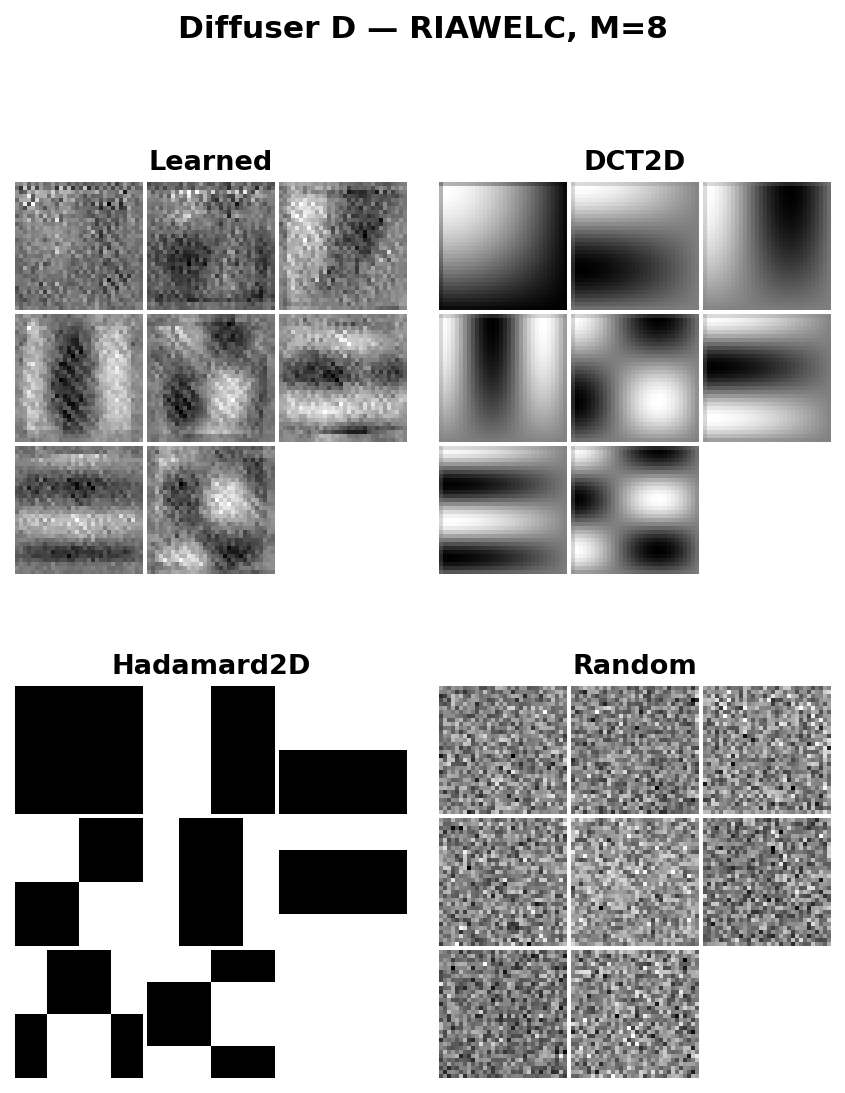}
  \caption{RIAWELC, $M{=}8$, all four modes}
\end{subfigure}\hfill
\begin{subfigure}{0.30\linewidth}
  \includegraphics[width=\linewidth]{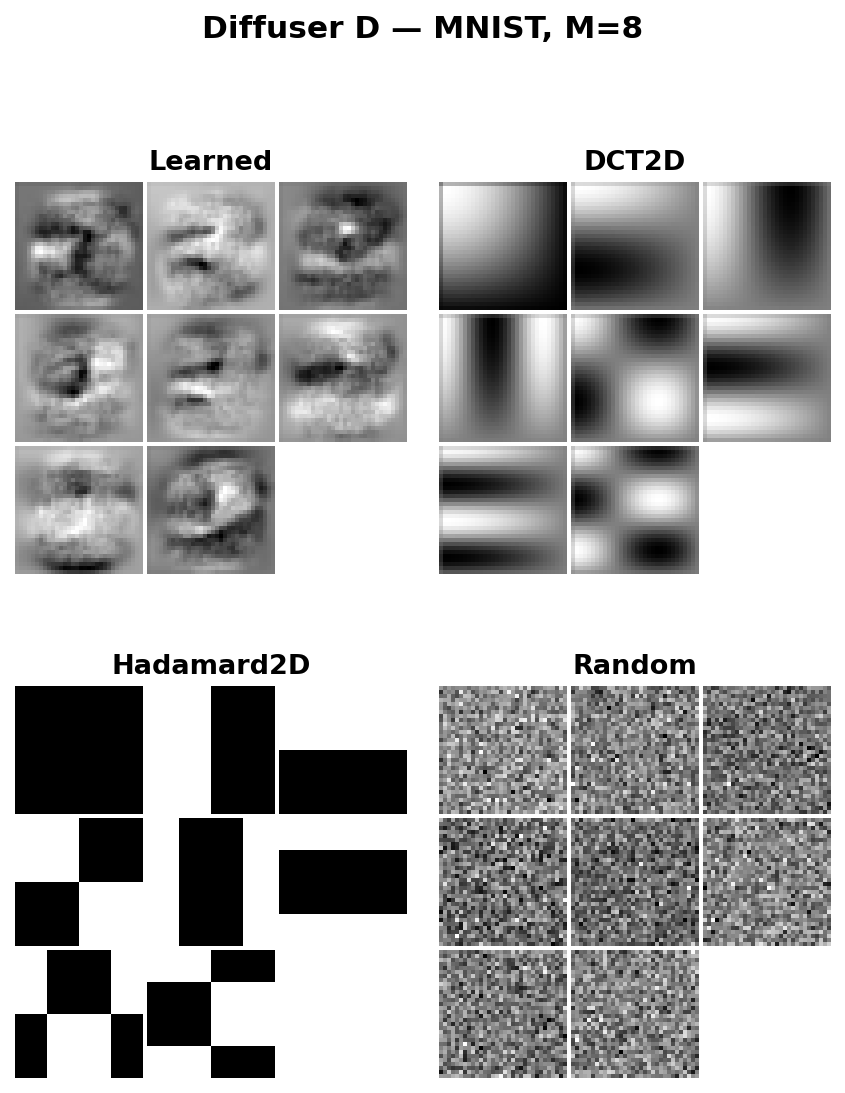}
  \caption{MNIST, $M{=}8$, all four modes}
\end{subfigure}\hfill
\begin{subfigure}{0.30\linewidth}
  \includegraphics[width=\linewidth]{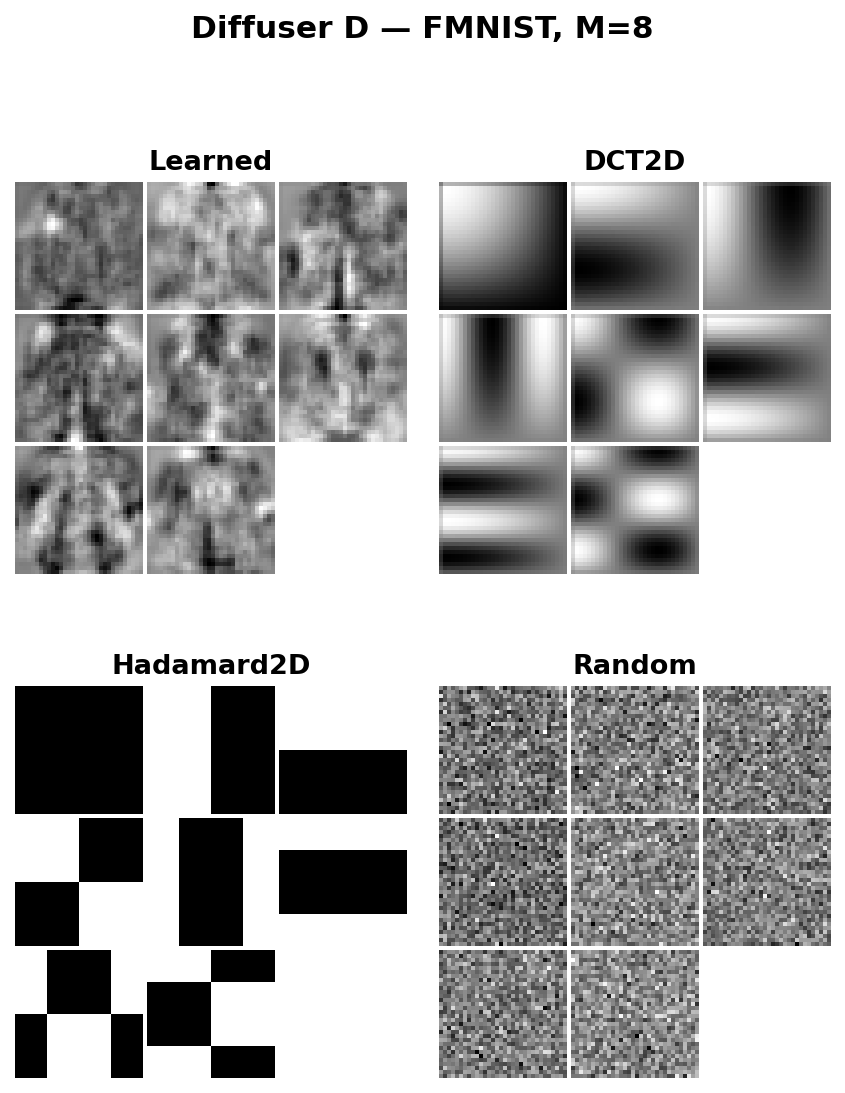}
  \caption{F-MNIST, $M{=}8$, all four modes}
\end{subfigure}

\vspace{0.5em}
\begin{subfigure}{0.23\linewidth}
  \includegraphics[width=\linewidth]{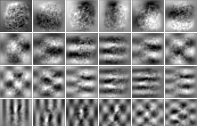}
  \caption{MNIST, $M{=}128$,\\learned (24/128 rows)}
\end{subfigure}\hfill
\begin{subfigure}{0.23\linewidth}
  \includegraphics[width=\linewidth]{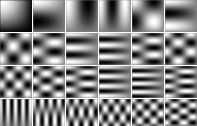}
  \caption{MNIST, $M{=}128$,\\DCT2D (24/128 rows)}
\end{subfigure}\hfill
\begin{subfigure}{0.23\linewidth}
  \includegraphics[width=\linewidth]{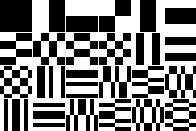}
  \caption{MNIST, $M{=}128$,\\Hadamard2D (24/128 rows)}
\end{subfigure}\hfill
\begin{subfigure}{0.23\linewidth}
  \includegraphics[width=\linewidth]{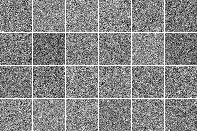}
  \caption{MNIST, $M{=}128$,\\Random (24/128 rows)}
\end{subfigure}
\caption{Learned pattern set against all three fixed transforms, each row of
$\D$ reshaped to $32\times32$ and tiled. Top row, $M=8$ (all 8 rows, every
mode): the learned rows (top-left quadrant of each panel) carry
content-related structure that DCT2D, Hadamard2D, and Random do not,
including on the target application, RIAWELC. Bottom row, $M=128$: at this
budget all 128 rows compressed onto the page are illegible, so we show the
first 24 of 128 rows for every mode (row order is storage order, not sorted
by appearance, so this is a representative crop, not a selected one; code
for the full $128$-row mosaics is part of the repository linked in
Section~\ref{sec:recipe}). Structurally, the learned rows have visually
converged onto their DCT2D anchor by $M=128$ and remain visually distinct
from the blockwise Hadamard2D basis and from unstructured Random, even
though by this budget all four modes reach comparable accuracy
(Table~\ref{tab:main}); the structural resemblance is to the specific
anchor the correction was initialized from, not a general convergence
across every fixed transform.}
\label{fig:masks}
\end{figure*}

Rendering the $M$ rows of $\D$ as $32\times32$ tiles (Fig.~\ref{fig:masks}),
the low-budget learned rows carry content-related texture on RIAWELC, with
faint horizontal banding in several tiles, centrally concentrated strokes on
MNIST, and bilaterally symmetric garment-like silhouettes on Fashion-MNIST,
visibly different from the three fixed transforms shown alongside them in
the same panels. We show the target application, RIAWELC, alongside the
digit benchmarks here rather than describing it in prose only, so the reader
can judge the pattern directly. By $M=128$ the learned rows visually
resemble the DCT2D initialization they are corrected around (Fig.~\ref{fig:masks},
bottom row), where the measured learned-versus-DCT2D margin is small, while
remaining visually distinct from the Hadamard2D and Random patterns shown
alongside it at the same budget. This is consistent with
the intuition that the correction $\bDelta$ is important especially when the
budget is tight. (We did not repeat on RIAWELC the quantitative
symmetry/centering structure analysis carried out for the digit benchmarks
elsewhere; the description above is a visual reading, not a measured
statistic.)

\subsection{Ablations}
\label{sec:ablations}

\noindent\textbf{Anchor: DCT2D versus Hadamard2D.}
DCT2D is the best fixed baseline; we test whether it is also the best
\emph{anchor} for $\bDelta$. Replacing $\Dbase$ with a sequency-ordered
Hadamard2D basis (Table~\ref{tab:anchor}) gives a genuinely mixed result. On
RIAWELC the DCT2D anchor wins at all five budgets, by a small
one-directional margin. On MNIST the Hadamard2D anchor wins or ties at four
of five budgets. On Fashion-MNIST neither direction is consistent. Most of
the gaps are within the project's run-to-run noise, roughly $0.8$ to $1.6$
points, so we do not claim DCT2D is a universally better anchor, only that
it is a safe default and, separately, the best \emph{fixed} baseline.

\begin{table}[t]
\centering
\caption{Learned residual accuracy (\%) by anchor basis.}
\label{tab:anchor}
\small
\begin{tabular}{@{}ll rrrrr@{}}
\toprule
Dataset & Anchor & $M{=}8$ & $16$ & $32$ & $64$ & $128$ \\
\midrule
\multirow{2}{*}{RIAWELC}
 & DCT2D & 81.50 & 85.80 & 86.28 & 86.32 & 85.82 \\
 & Hadam. & 79.68 & 85.23 & 85.41 & 85.14 & 85.59 \\
\midrule
\multirow{2}{*}{MNIST}
 & DCT2D & 95.27 & 97.81 & 98.80 & 98.98 & 98.75 \\
 & Hadam. & 95.48 & 98.12 & 98.73 & 98.99 & 98.93 \\
\midrule
\multirow{2}{*}{F-MNIST}
 & DCT2D & 84.50 & 87.92 & 88.27 & 88.82 & 88.95 \\
 & Hadam. & 84.93 & 87.74 & 88.70 & 89.04 & 88.70 \\
\bottomrule
\end{tabular}
\end{table}

\medskip\noindent\textbf{Free versus network-generated $\bDelta$ on RIAWELC.}
The network-generated correction keeps the large low-budget advantage,
$+10.17$ points at $M=8$ and $+3.14$ points at $M=16$, and it substantially
narrows the high-budget deficit: $-1.12$ and $-0.34$ points at $M=64$ and
$M=128$, against $-3.14$ and $-3.51$ points for a free $M\times D$ tensor
under the same decoder. Constraining the correction's degrees of freedom
helps where the free tensor was overfitting, which supports the
small-training-set explanation for treating RIAWELC differently.

\section{Discussion and Limitations}
\label{sec:limitations}

\subsection{Why the low-budget concentration is the useful part}
In dose- or time-constrained imaging, $M$ sets the physical cost directly. A
practitioner who already classifies on compressed single-pixel measurements
can replace a fixed transform with a jointly trained learned diffuser,
without changing the classifier, and gain accuracy at exactly the small-$M$
operating point. Equivalently, a target accuracy can be reached at a smaller
$M$, and therefore potentially at a lower delivered dose or a shorter acquisition. The
same argument should apply also to clinical X-ray and CT, where
$M$ maps to patient dose, However, we leave this avenue for future research. The approach is believed to be compatible with existing hardware: the learned operator is still a single
$M\times D$ matrix, so a suitably programmable acquisition pipeline may need no new
hardware, only a different pattern set loaded onto the modulator.

\subsection{Limitations}
\begin{itemize}
\item \textbf{Single run per cell.} There is no multi-seed average. Any
result that is not backed by a clear one-directional trend across $M$, for
example parts of the anchor ablation and the exact location of the RIAWELC
crossover, should be read as provisional. Multi-seed replication is the
first item of future work.
\item \textbf{RIAWELC split.} The $80/20$ split is at the crop level.
Multi-class welds rule out the simplest label-leakage shortcut, in which
weld identity would predict the class, but a residual risk of near-duplicate
crops on either side of the split remains. A weld-level split was not run
under the current pipeline.
\item \textbf{Simulation only.} The measurement $\y=\D\x$ is computed
numerically. There is no optical bench and no detector-noise model, so shot
noise, read noise, and pattern-projection error are all absent. Robustness
of the learned patterns under realistic noise is untested, and is a natural
next experiment since noise could plausibly erode a finely tuned $\bDelta$
faster than it erodes a generic basis.
\item \textbf{RIAWELC uses a different decoder.} The adjoint-proxy
convolutional decoder for RIAWELC differs from the multilayer perceptron
used for the digit benchmarks, so absolute accuracies are not comparable
across datasets. Every claim in this paper is scoped within a single
dataset; only the direction and rough size of the learned-versus-fixed
margin is compared across datasets.
\item \textbf{Physical realizability is only partial.} Every pattern set
here, fixed and learned alike, has signed entries and full
floating-point precision. A real digital micromirror device or spatial light
modulator needs non-negative or dual-rail sign encoding and low bit depth.
These costs fall equally on the fixed baselines and on the learned operator,
so they do not change the comparison, but they must be addressed before any
hardware claim, and quantization in particular may interact with the small
learned corrections.
\end{itemize}

\section{Conclusion}
\label{sec:conclusion}
In LIPS-PC, learning the measurement operator end-to-end for the classification task,
under the constraint that it stay one fixed linear matrix a single-pixel
apparatus could apply, gives a large and consistent accuracy advantage over
every fixed transform at the smallest budget: $9$ to $15$ percentage points
at $M=8$ across a real industrial X-ray dataset and two standard benchmarks,
equivalently a $2\times$ to $4\times$ saving in physical measurements at
matched accuracy. The advantage generally decreases as the budget grows,
towards zero on the digit benchmarks and to a small reversal on RIAWELC at
intermediate budgets, consistent with the task-driven-embedding view
predicts: the choice of which few projections to take matters most when
there are very few to take. The practical statement is therefore a scoped
one. A task-learned linear operator is a candidate replacement for a fixed
transform that is worth its training cost specifically in the dose- and
time-limited, sub-Nyquist regime, and not a general-purpose substitute for
classical bases. Sharpening the result would mean multi-seed replication
(the current numbers are single-run), a weld-level RIAWELC split, a
detector-noise model, and sign- and precision-constrained pattern sets as a
bridge toward an optical or X-ray bench.

\section*{Acknowledgment}
The authors thank Amotz Taub-Tabib for helpful discussions. E.C. was supported by the European Union's Horizon Europe research and innovation programme under grant agreement No. 101178170, by the Israel Science Foundation under grant agreement No. 2208/24, by the Israel Innovation Authority, and by the Israeli Council for Higher Education through ``QERNEL'' and the ``Interdisciplinary Center for the Theory of Quantum Computing''.


\end{document}